\documentclass[galaxies,article,accept,pdftex,moreauthors]{Definitions/mdpi} 

\firstpage{1} 
\pubvolume{11}
\issuenum{1}
\articlenumber{32}
\pubyear{2023}
\copyrightyear{2023}
\externaleditor{Academic Editor: Christian Corda}
\datereceived{15 November 2022} 
\daterevised{8 December 2022} 
\dateaccepted{19 January 2023} 
\datepublished{15 February 2023} 
\hreflink{https://doi.org/\linebreak 10.3390/galaxies11010032} 

\usepackage[utf8]{inputenc}

\usepackage{tikz} 					
\usetikzlibrary[patterns]				
\usetikzlibrary[arrows]				
\usetikzlibrary{positioning}			
\usepackage{marginnote}
\usepackage{framed}			

\Title{The Next Generation Event Horizon Telescope Collaboration: History, Philosophy, and Culture}

\TitleCitation{The Next Generation Event Horizon Telescope Collaboration: History, Philosophy, and Culture}

\Author{
	Peter Galison  $^{1,2,3,}$*, 
	Juliusz Doboszewski $^{1,4,}$*,
	Jamee Elder $^{1,4,}$*\orcidA{},
	Niels C. M. {Martens} $^{5,4,6,7,}$*\orcidB{}, 
	Abhay Ashtekar $^{8}$,
	Jonas~Enander {$^{9}$,}  
	Marie Gueguen $^{10}$\orcidC{},
	Elizabeth A. {Kessler} $^{11}$,
	Roberto Lalli $^{12,13}$,
	Martin Lesourd $^{1}$,
	Alexandru~Marcoci~$^{14}$\orcidD{}, 
	Sebastián {Murgueitio Ramírez} $^{15}$\orcidE{},
	Priyamvada Natarajan $^{1,16,17}$,
	James Nguyen $^{18}$,
	Luis~Reyes-Galindo $^{19}$\orcidF{},
	Sophie Ritson $^{20}$\orcidG{},
	Mike D. {Schneider} $^{21}$,
	Emilie Skulberg $^{22,23}$,
	Helene Sorgner $^{7,24}$,
	Matthew~Stanley $^{25}$,
	Ann C. {Thresher} $^{26}$, 
	Jeroen Van Dongen $^{22,23}$, 
	James Owen {Weatherall} $^{27}$,
	Jingyi {Wu} $^{27}$ 
	and~Adrian~W\"uthrich $^{7,28}$ \orcidH{}
}

\AuthorNames{
	Peter Galison
	Juliusz Doboszewski
	Jamee Elder
	Niels C. M. {Martens}
	Abhay Ashtekar
	Jonas Enander
	Marie Gueguen
	Elizabeth A. {Kessler}
	Roberto Lalli
	Martin Lesourd
	Alexandru Marcoci
	Sebastián {Murgueitio Ramírez}
	Priyamvada Natarajan
	James Nguyen
	Luis Reyes-Galindo
	Sophie Ritson
	Mike D. {Schneider}
	Emilie Skulberg
	Helene Sorgner
	Matthew Stanley
	Ann C. {Thresher}
	Jeroen {van Dongen}
	James Owen {Weatherall}
	Jingyi {Wu}
	Adrian W\"uthrich}

\AuthorCitation{{Galison, P.; Doboszewski, J.; Elder, J.; Martens, N.C.M.; Ashtekar, A.; Enander, J.; Gueguen, M.; Kessler, E.A.; Lalli, R.; Lesourd, M.; et al.
}}  

\address{%
	$^{1}$ \quad Black Hole Initiative, Harvard University, Cambridge, MA 02138, USA\\ 
	$^{2}$ \quad Department of History of Science, Harvard University, Cambridge, MA 02138, USA\\
	$^{3}$ \quad Department of Physics, Harvard University, Cambridge, MA 02138, USA\\
	$^{4}$ \quad Lichtenberg Group for History and Philosophy of Physics, University of Bonn, 53113 Bonn, Germany\\
	$^{5}$ \quad Freudenthal Institute, Utrecht University, 3584 CC Utrecht, The Netherlands\\
	$^{6}$ \quad Institute for Theoretical Particle Physics and Cosmology, RWTH Aachen University, 52074 Aachen, Germany\\
	$^{7}$ \quad Epistemology of the LHC Research Unit, {University of Wuppertal,} IZWT, Gaußstraße 20, \mbox{42119 Wuppertal, Germany} 
	\\
	$^{8}$ \quad Physics Department \& Institute for Gravitation and the Cosmos, Pennsylvania State University, State~College,~PA 16802, USA \\
    $^{9}$ \quad Independent scholar, Stockholm, Sweden\\
	$^{10}$\quad Institute of Physics, University of Rennes 1, 35042 Rennes, France\\ 
	$^{11}$\quad American Studies Program, Stanford University, Stanford, CA 94305, USA \\
	$^{12}$\quad Max Planck Institute for the History of Science, 14195 Berlin, Germany\\
	$^{13}$\quad {Dipartimento di Ingegneria Meccanica e Aerospaziale}, 
	Politecnico di Torino, 10129 Turin, Italy\\
	$^{14}$\quad {Centre for the Study of Existential Risk, University of Cambridge, Cambridge CB2 1SB, UK} \\ 
	$^{15}$\quad {Department of Philosophy}, Purdue University, West Lafayette, IN 47907, USA\\ 
	$^{16}$\quad Department of Astronomy, Yale University, New Haven, CT {06511}, USA\\ 
	$^{17}$\quad Department of Physics, Yale University, New Haven, CT 06520, USA\\
	$^{18}$\quad Department of Philosophy, Stockholm University, 10691 Stockholm, Sweden\\
	$^{19}$\quad Knowledge Technology and Innovation/Philosophy Groups, Wageningen University, 6708~WG~Wageningen,~The Netherlands\\
	$^{20}$\quad Research Innovation Commercialisation, The University of Melbourne, Parkville 3010, Australia\\
	$^{21}$\quad Department of Philosophy, University of Missouri, Columbia, MO {65201}, USA\\
	$^{22}$\quad Institute of Physics, University of Amsterdam, 1090 GL Amsterdam, The Netherlands\\
	$^{23}$\quad Vossius Center for History of Humanities and Sciences, University of Amsterdam, 1090~GL~Amsterdam,~The~Netherlands\\
	$^{24}$\quad Department of Science Communication and Higher Education Research \&  Department of Science,  Technology~and Society Studies, University of Klagenfurt, 9020 Klagenfurt,  Austria\\
	$^{25}$\quad Gallatin School of Individualized Study, New York University, New York, NY 10003, USA\\
	$^{26}$\quad McCoy Family Center for Ethics in Society {and the Stanford Doerr School for Sustainability}, Stanford~University,  Stanford, CA 94305, USA\\ 
	$^{27}$\quad Department of Logic and Philosophy of Science, University of California, Irvine, CA 92697, USA\\
	$^{28}$\quad Institute of History and Philosophy of Science, Technology, and Literature, Technische Universität Berlin, 10623 Berlin, Germany

}

\corres{Correspondence: galison@fas.harvard.edu (P.G.); jdobosze@uni-bonn.de (J.D.); jelder@fas.harvard.edu~(J.E.);  n.c.m.martens@uu.nl (N.C.M.M.)
}

\abstract{This white paper outlines the plans of the History Philosophy Culture Working Group of the Next Generation Event Horizon Telescope Collaboration.}


\begin{document}


	
	
	
	
	
	\section{Introduction}

	\textls[-5]{\textbf{Coordinating author: Galison, P.; Contributing authors: Elder, J. and Thresher,~A.C.}}

	{Deep in the} 
	development of physics lie crucial intersections of science and philosophy. 
	When Isaac Newton released his \emph{Principia Mathematica} 
	to the world, he included a ``Scholium'' on space and time. It contains no diagrams, mathematical expressions, experimental reports, theorems, or specific laws of motion or gravity. Instead, the Scholium sets out the starting terms of the inquiry itself, delving into the nature of space, time, and place. ``I must observe'', Newton insisted, ``that the common people conceive those quantities under no other notions but from the relation they bear to sensible objects. And thence arise certain prejudices, for the removing of which it will be convenient to distinguish them into absolute and relative, true and apparent, mathematical and common.'' Rulers and clocks, calendars and sunrises, all the motions we use to tell time: these were merely the observable, ``sensible'' aspects of our basic concepts. How, asked Newton, are we ``to obtain the true motions from their causes, effects, and apparent differences, and the converse.'' These deeply philosophical questions motivated the writing of the \emph{Principia} (\citep{newton1689}, pp. 6--12).
	
	For Einstein, too, philosophical analysis was essential to subverting conformist tendencies in approaching central questions of physics. Nowhere was this more important than in his relativity theories: first, in his revision of space, time, and simultaneity in special relativity, leading to the unified spacetime introduced by Hermann Minkowski; and second, in Einstein’s far deeper 1915 reconfiguration of spacetime in general relativity.\endnote{A very helpful framing of the history of general relativity can be found in~\citep{eisenstaedt2006}. On Einstein’s special theory of relativity, focusing on his redefinition of simultaneity, see~\citep{galison2003}. On the eclipse expedition of 1919 and its surround---as a historical example of observational history, see~\citep{kennefick2019,stanley2019}. On Einstein’s own trajectory to general relativity, see~\citep{renn2005}.} 
	Einstein drew on a range of philosophical influences: from his youth forward, Einstein maintained a {persisting} 
	interest in the work of Immanuel Kant and the neo-Kantians; he and his ``Olympia Academy'' dug line-by-line into Henri Poincaré’s work on conventionalism; he sustained an abiding, if critical, interest in the work of the Vienna Circle; he also borrowed from Ernst Mach, who was deeply suspicious of an absolute, sense-independent notion of space and time. 
	Throughout his life, Einstein believed that epistemology---the study of the formation, nature and justification of knowledge---and science ``are dependent upon each other. Epistemology without contact with science becomes an empty scheme. Science without epistemology is---insofar as it is thinkable at all---primitive and muddled.'' (\citep{einsteinremarks}, pp. 683--684).\endnote{On the philosophically-inflected work of Einstein, see, as an entr\'{e}e into the literature~\citep{holton1988,ryckman2005,vandongen2010,janssen2014,
			howardgiovanelli2019}; and for a launch into the philosophy in Einstein’s physics see~\citep{ryckman2005,norton2019}. Of books on the philosophy of spacetime, Earman’s have been a grounding point of many discussions~\citep{earman1992,Earman1995bcwa}, as has the (physics-based) lapidary take on general relativity by Wald~\citep{wald1992}. For a fine example of a more recent conceptual analysis, see~\citep{curiel2019}.} 
	
	We are now in a golden age of astronomy replete with extraordinary astrophysical objects. Of these, none has elicited as much fascination as black holes. The one-way membrane of the event horizon, the inner region where spacetime trajectories can cross themselves, and the singular breakdown of spacetime structure are just some of the provocations that black holes have presented to history and philosophy of science. Black holes thus present an opportunity to continue the tradition of intertwining groundbreaking physics with historical, philosophical, and {cultural} analysis.    
	
	From its start in 2015/16, the Black Hole Initiative (BHI) has set the history and philosophy of black holes alongside mathematics, physics, and astronomy as a crucial disciplinary ingredient.\endnote{On the long-term history of relativity as it opened up into the science of black holes in particular, see~\citep{thorne1995}.} Though based at Harvard, the BHI has  drawn on collaborators far beyond its halls.  
	Many of the scientists within the BHI have also been involved with the Event Horizon Telescope (EHT), a long-running, planetary-scale virtual telescope composed of widely-dispersed observatories. 
	
	The EHT observatories, eight on six sites as of 2017, and expanded to eleven observatories since, register millimeter electromagnetic waves from the same source by putting the data on hard drives with precise time stamps given by a hydrogen maser. The drives are then transported to central computing facilities where supercomputer ``correlators'' align the recorded signals. These aligned data can then be used to create images. 
	In April 2019, the EHT Collaboration released the first ever picture of a black hole, M87*, the 6.5 billion solar mass compact object at the center of the elliptical galaxy M87 in the constellation Virgo~\cite{2019EHTpaper4}. Three years later, the EHT issued an image of the supermassive black hole, Sgr~A*, at the center of the Milky Way~\cite{EHT2022_paper3}. Extending this work, the next generation EHT (ngEHT) aims to supplement the EHT network of telescopes with an additional ten or more sites that would fill out the virtual telescope and bring in new hardware and software, that together would make possible higher-resolution pictures and even movies.\endnote{See (\cite{EHT2022_paper3}, Sections 4.4 and 9) for discussion of ``dynamic imaging'', which results in a movie of the source (i.e., a series of images or frames) instead of a single image.}
	
	In the imaging campaign leading to the first pictures of M87* and Sgr~A*, cross-fertilisation of science studies with the work of the EHT imaging group placed black hole images within a broader historical-epistemic context of pictorial argumentation. This allowed the objectivity of the black hole images to be framed in terms of longer-term and analytic approaches to the objectivity of images~\citep{dastongalisonobjectivity}. 
	The goal now is to expand this imbrication in the next generation Event Horizon Telescope, setting the History, Philosophy, and Culture (HPC) Working Group as one of the eight science working groups of the collaboration as of 2022. These working groups will bring to bear on the study of black holes the resources of the history and philosophy of science along with the panoply of disciplines that compose Science and Technology Studies (STS). More specifically, the goal is to put this interdisciplinary working group into productive conversation with the other science and technical working groups---in the process of research and not as a post hoc account. Parallel to the other working groups, HPC will divide into four focus groups: 
	\begin{enumerate}
		\item Algorithms, Inference, and Visualization,
		\item Foundations,
		\item Collaborations,
		\item Siting, Education, In- and Outreach.
	\end{enumerate}
	
	The Algorithms, Inference and Visualization (AIV) focus group aims to understand the epistemic and aesthetic choices that will guide ngEHT image production. To do so, the group will work closely with the Algorithms and Inference Working Group of the ngEHT. The AIV focus group provides a philosophical, historical, and social scientific complement to this working group, providing a space for a comparative discussion of inference methods and the broader social context of image dissemination. 
	In this article (Section \ref{AIV}) we will report on the power and limits of ``robustness'' as an analytic virtue, and on the visual conventions of the EHT and ngEHT to come. 
	
	The Foundations focus group builds on the existing BHI Foundations Seminar, which draws historians, philosophers, and scientists to its meetings on topics ranging from the thermodynamics of black holes to the nature of singularities.  
	In this article (Section \ref{foundations}) we discuss the relationship between theory and observation, through selected topics of foundational interest (e.g., no hair theorems) that illustrate the often-complex nature of this~relationship.  
	
	Alongside these bridges between history, philosophy and scientific work are questions about the constitution of the ngEHT. What structure should its governance have? How should the collaboration ensure transparency, choose scientific goals, and assure representation in decision-making? What rules of the road should guide comportment in the collaboration, ranging from authorship and credit to collegiality, diversity, equity, and inclusion? Such questions will be addressed by the Collaborations focus group. Here we include a preliminary discussion of these issues (Section \ref{collaboration}), drawing not only on the History and Philosophy of Science (HPS) but on the broader mix of Science and Technology Studies (STS) (including sociological and ethnographic work). To these questions, we offer initial reflections on the broad range of topics within the purview of the AIV, Foundations, and Collaborations focus groups---\emph{initial, not final}, as befits these early, formative days of the ngEHT.

	One important note: we acknowledge the cultural, historical, epistemic, political, environmental, and economic issues that surround the siting of telescopes. These problems have recently been at the fore of both academic and public interest due to ongoing conflicts at places like the Thirty Meter Telescope in Hawai'i, and the Square Kilometre Array in South Africa {and Australia}
 , where local communities have protested the projects for reasons including a lack of inclusion, concern for religious, cultural, and environmental sites, and the ongoing role of science within the longer history of colonialism and self-determination.\endnote{Two excellent doctoral dissertations offer fine-grained analysis of the mountaintop dispute, and include a wide range of further references.~\citet{SwannerThesis} focuses on the triply conflicting astronomical, environmental and indigenous narratives that collided at Mt. Graham, Mauna Kea, and Kitt Peak;~\citet{SalazarThesis} addresses the Kanaka rights claim, specifically addressing the Thirty Meter Telescope (TMT), in opposition to a framing of the dispute as one of ``stakeholders'' or a ``multicultural'' ideal. Swanner focuses on Mauna Kea in a subsequent article, also on the TMT~\citep{Swanner2017}. For an important current Hawaiian-led impact assessment of the TMT including additional references, see Kahanamoku et al. \cite{Kahanamoku2022}. Many further references across a wide cross-disciplinary range including archaeology, biology, among others, will be given in a subsequent paper directed toward siting.} These sites, and others, highlight the need for careful discussions of our ethical obligations towards local communities, individuals, and the environment when building instruments.\endnote{Highlighting the environmental, social, experimental, and ethical implications of locating scientific facilities through a robust history of locating LIGO’s sites, see Nichols, T. \cite{Nichols2022,Nicholsnd}} Given the importance of such topics, we have decided more serious work is required before we comment on the normative aspects of siting. As such, we will not be discussing siting in this paper, but are instead determined to build and maintain a broadly-diverse, appropriately interdisciplinary focus group dedicated to the topic, drawing on community members, scientists, philosophers, humanists, and social scientists to frame these issues. We anticipate producing publications dedicated solely to this topic in the near future.
	
	\section{Algorithms, Inference, and Visualization} \label{AIV}

\textbf{Coordinating author: Doboszewski, J.; Contributing authors: Elder, J.; Enander, J.; Galison, P.; Gueguen, M.; Kessler, E.A.; Nguyen, J.; Skulberg, E.; Stanley, M. and Van Dongen, J.}

	
	\subsection{Introduction}\label{aiv_intro}
	
	The Algorithms, Inference, and Visualization (AIV) focus group is a space for a general and comparative discussion of inference methods. The overarching goal is to analyse (and  also contribute to) the epistemic and aesthetic choices that will guide ngEHT image production and interpretation. Many lessons can be learned from other computationally heavy areas of science (such as climate science or cosmological simulations) and other large experiments in physics. Here we discuss two example clusters of questions of interest to the AIV: robustness and reliability of imaging methods, and aesthetic choices in black hole imaging. 
	A broader look at such issues will allow us to keep track of the range of factors contributing to decision-making, leading to better-informed choices in the long run.
	
	\subsection{Robustness and Reliability of Imaging}\label{robustness}
	
	``Robustness'' is  often used in discussions of EHT data and results, including the analyses of both M87* and Sgr~A*. Here we offer a short guide to its different uses in the scientific and philosophical literature, before we turn to discussing the use of robustness in justifying EHT and ngEHT results. 
	
	The robustness of a result can be characterized as the claim that if a variety of derivations, tests, or lines of evidence converge on a result, then that result is more secure than if it were obtained with only a single line of evidence. For that boost in confidence to hold, lines of evidence should be, in some sense, independent: convergence should not be attributable to some mistaken or irrelevant assumption shared by all lines of evidence (although see~\citep{schupbach2018robustness} for a discussion of the difficulty explicating what this amounts to).\endnote{If ``secure'' is understood in terms of degrees of belief (expressed by some function satisfying the Kolmogorov axioms of probability), then ``boost in confidence'' can be understood as (something like) the statement that the conjunction $E_{1} \& \ldots \& E_{n}$ confirms $R$ to a greater extent than $E_i$ alone, for any $i$; where $R$ is the result, and $E_{i}$ are lines of evidence.}
	
	Experimental results are robust in the above sense when aspects of the experimental setup are varied, but results nonetheless converge---for example, when multiple independent measurements of  Avogadro's number produce consistent results, these results are considered to be robust. In typical experimental situations, many factors can be varied, including the sample population or control group, initial or boundary conditions, and the measurement apparatus. Many such variations are impossible in the (ng)EHT, which  will deal with a small number of sources, initially sparse sampling, lack of control over sources, and a lack of alternative instruments capable of performing the same measurements.
	However, multiple redundancies are built into the EHT measurements. For example, the use of varied calibration pipelines builds confidence that the result is not due to idiosyncratic factors in a particular pipeline.
	For some purposes like mass measurements other means of accessing the system (e.g., observations of S stars orbiting Sgr~A*) also contribute to the robustness of the EHT results. 
	
	The results of modeling and data analysis methods can also be called robust when they are consistent across variations in modeling assumptions, analysis methods, or parameter choices. 
	The robust occurrence of some features (e.g., the temperature increase for a range of climate simulations, or ring size for a range of EHT imaging methods) increases confidence in that aspect of the modeling outcomes, while other, less stable features (e.g., regional precipitation for climate simulations, the positions of bright `knot' structures in EHT images of Sgr~A*) should be treated with caution. 
	
	
	Among the main lines of criticism formulated against robustness arguments, two seem especially relevant in the context of the ngEHT. 
	Such criticisms envisage the ensemble of models containing (1) a shared core of assumptions, which make the models comparable, and (2) an unshared part, deemed problematic (e.g., modeling assumptions, idealizations, parametrizations or {measurement} apparatus), whose possible impact on the {models'} output must be understood and eliminated.

	The first criticism argues that in numerical models the shared core common to all models tends to include problematic assumptions. Common idealizations, such as iterative and discretization errors, are unavoidable to numerically solve the problem but are also important sources of numerical artifacts. Hence, their impact cannot be determined through robustness reasoning. 
	The second criticism points out that the mere convergence of results cannot by itself indicate a reliability or partial truth: something else is needed. \mbox{\citet{Gueguen2020}} examines a number of cases where convergent results across N-body simulations may be attributable to numerical artifacts. For example,~\citet{Baushev-etal2017} point out that N-body cosmological simulations predict a ``cuspy'' profile for dark matter halo density for galaxy center regions (in conflict with observations). They argue that the convergence of simulations on such predictions is produced by numerical artifacts rather than by a physically realistic process captured by the simulations. This case shows how the apparent robustness of simulation results may not indicate that the results are reliable.  
	As emphasized by~\cite{Orzack-Sober1993} in their response to the seminal paper by~\cite{Levins1966} on robustness, from a purely logical point of view, robustness can guarantee reliability only in those cases where we already know that one of the  models in the set is correct.\endnote{Here we retain Orzack and Sober's terminology, describing models as ``true'' or ``correct''. Note, however, that this terminology is controversial (see Section \ref{theory_observation}) with some recent philosophical treatments of models suggesting that models themselves are neither true nor false.} This condition is rarely satisfied when robustness is the most needed, i.e, when it is used to supplement the absence of analytic solutions or experimental measures that could determine whether one of the models is indeed correct. Hence there is a clear need to analyze when robustness is an efficient tracer of reliability within the ngEHT program, and when it needs to be supplemented or substituted. 
	
	In the suite of papers that the EHT issued on M87*~\cite{2019EHTpaper1,2019EHTpaper2,2019EHTpaper3,2019EHTpaper4,2019EHTpaper5,2019EHTpaper6,2021EHTpaper7,2021EHTpaper8} and Sgr~A*~\cite{EHT2022_paper1,EHT2022_paper2,EHT2022_paper3,EHT2022_paper4,EHT2022_paper5,EHT2022_paper6}, the collaboration's overwhelming concern was to establish, with confidence, the existence of a ring surrounding the black hole shadow. That is, the EHT Collaboration did not want to issue a false positive. For that reason, in the M87* image work \textit{robustness} was key; the collaboration: varied the priors to make sure those choices were not forcing the image to be a ring; isolated four image-making groups to avoid cross-contaminating expectations based on others' results; and varied image reconstruction methods to ensure that the observed ring was not an artifact of any one imaging method. These measures constituted a determined drive to be sure that in the image of M87* the ring and bright crescent in the south were as unshakeable as possible. 
	The commitment to robustness came with an unavoidable cost: other, valid effects---observations outside the ring, for example---might have been omitted. However, especially for this first, momentous publication, the collective desire was for an appropriately robust, and therefore conservative, claim.
	
	Yet, robustness is not the only possible epistemic desideratum. Over the course of the next generation of work, we may well want to pursue other, complementary ambitions. With highly specific models, physicists and observers could explore other predicted phenomena that might otherwise be lost in the noise. More unsteady, delicate phenomena in the accretion disk and jet formation, for example, could be detected using models and templates of various kinds. In particle physics, such targeted searches are common---this is what triggers do when they pluck a particular signal, interaction, particle, or phenomenon out of the vast sea of other results. Indeed, in many domains of physics, initial statements of groundbreaking results are more statistically fragile.\endnote{On the contrast between inclusive and selective instrumental demonstration in particle physics, see~\citet{galison1997image}.} Robustness is thus a core epistemic {virtue}, 
    but not the only one: too strong an emphasis on it could lead to false negatives by blinding us to hard-to-see phenomena just above the noise. Selectivity, pushed too hard, can produce false positives, giving us back what we hope and expect to see. We need \textit{both} robustness \textit{and} model-based selectivity. However, there are epistemic trade-offs to be made between these different epistemic virtues. Future work with the ngEHT will involve decisions about which virtues to prioritize in which contexts. 
	
	\subsection{Science and Aesthetics in Black Hole Imaging}\label{visualization}
	
	All images have style expressed through ``shared visual features'' (\citep{kemp2000visualizations}, p. 4). Graphs, for instance, tend to avoid detail. Certain color schemes are more used than others. Including or removing artifacts is another choice. 
	Astronomical images, whether based on empirical data or simulations, reflect an array of choices and decisions, and they also participate in their larger historical and cultural contexts. Their creation and interpretation rely on pre-existing visual traditions that establish the norms, expectations, and methods by which a scientific image is given meaning. The AIV focus group will draw on the extensive scholarship on imaging in astronomy and physics to reflect on such image-making choices and decisions by the ngEHT, as well as how the results are received and understood both within the scientific community and beyond~\citep{Bigg:2015aa,Bigg:2017aa,dastongalisonobjectivity, elkins2008six,fineman2019apollo, galison1997image,Hentschel:2000aa,Hentschel:2002aa,kaiser2005drawing,lane2010geographies,Messeri:2016aa,Nall:2019aa,Nasim2011,Nasim:2013aa,Schaffer,pang1997stars,Stanley2022,tai2019,vertesi2021seeing}. 
	
	Over the last several decades, images have furthered scientific understandings of black holes. However, until the EHT images, these representations were based on astronomers’ calculations and simulations rather than observations. 
	In the early 1970s, researchers visualized the basic outlines of black holes but their images were still in a schematic style~\citep{Godfrey:1970aa,Bardeen,Cunningham:1973aa}. Later in the same decade, more detailed and naturalistic visualizations of black holes emerged: a film clip  by Leigh Palmer, Maurice Pryce, and William Unruh (unpublished, but shared in multiple lectures) and Jean-Pierre Luminet's black and white drawing of a black hole accretion disk~\citep{Luminet1979}. Yet later, color visualizations, such as those by Heino Falcke, Fulvio Melia, and Eric Agol of Sgr~A*, theorized how the black hole shadow might look if observed using VLBI~\citep{Falcke:2000aa} (see also~\citep{Fukue:1988aa} for the first visualizations in color).
	
	Simulations remained a critical part of the EHT imaging process, resulting in observations that integrated theory in interesting ways. New data imaging pipelines were developed and used together with a library of synthetic images produced by general relativistic magnetohydrodynamic simulations and general relativistic ray tracing~\citep{2019EHTpaper5, 2021EHTpaper8, EHT2022_paper5}. Comparing the observations with theoretical simulations was key for establishing that the observed ring was created by synchrotron emission from a hot plasma orbiting near the black hole.
	Although these specific techniques were novel, astronomers have long been aware of the dependence of their observations on theory. The need to reduce collected data to a more concise and tractable form in order to account for phenomena such as stellar aberration, atmospheric refraction, or the so-called ``personal equation'' (variations due to a specific observer's idiosyncrasies) means that astronomy as a discipline has reflected on the role of theory in making raw data into useful depictions of celestial bodies for generations~\citep{schaffer1988personaleq, stanley2013gitelman}. There is a long intellectual ancestry of ever-more complex reliance on theory to allow for increasingly powerful forms of observation and imaging. These increases in scope and depth, however, also required more delicate conceptual and social scaffolding, increasing the possible influence of bias and blind-spots~\citep{shapere1982obs,pinch1985obs}.  
	The EHT Imaging Group was keenly aware of concerns about bias and systematic error; from the beginning, the imaging process was shaped by these concerns, in order to ensure the validity of the image~\citep{2019EHTpaper4}. 
	
	\textls[-7]{Another concern for the EHT was the legibility of their images for a wide audience---particularly for the first image of M87*, given {its} novelty.} 
    The color palette---a ring in orange-red hues against a black background---was chosen with this in mind; orange was believed more likely to signify heat than blue (even though blue has shorter wavelengths and is therefore ``hotter'' than orange). Because the EHT Collaboration wanted to share one image with audiences of varying degrees of specialization (see~\citep{skulberg2021}, on the basis of~\citep{SaraRadboud}), a single averaged image was created from multiple images based on different imaging methods. Notably, the averaging of the Sgr~A* image was different than that of M87*, with the former averaging process being more complex than the latter (see~\citep{2019EHTpaper4} for M87* and~\citep{EHT2022_paper3} for Sgr~A*). These averaging techniques also connect to historical practices going back to the very beginning of technology-assisted scientific images with Galileo, Hooke, and Hevelius. Such figures used compositing techniques to make their early telescopic and microscopic images legible to wide, non-specialist audiences (particularly those who did not have access to the relevant instruments). Even through the nineteenth century, and well into the twentieth, it was accepted that astronomers would need to synthesize many individual observations in order to produce a reliable drawn or photographic image~\citep{pang1993socialevent,Nasim:2013aa,tai2017}. 
	
	Given that more than a billion people saw the M87* image within days of its release~\citep{Kessler:2019aa}, EHT imaging choices will continue to influence how black holes are perceived and understood. 
	The next generation of images produced by the ngEHT will build on these perceptions while introducing new considerations; increasing the bandwidth, including other frequencies, and adding telescope sites, will allow for greater resolution, and the production of moving images (movies). This means that further choices will need to be made about how to convey this information in an image.
	
	The history of astronomical images (and their reception) offers  models to consider. Many existing astronomical images use color to distinguish between different wavelengths, and the hues often signify both physical properties and evoke aesthetic responses. For example, color in many Hubble Space Telescope images indicates relative temperatures while also creating a resemblance to the sublime nineteenth-century paintings of the western regions of the USA~\citep{kessler2012picturing}. Such seemingly naturalistic color choices elicit questions from viewers, who assume color corresponds to human perception. In other instances—remote sensing of the Earth and some planetary images—more obviously engineered color choices enhance morphology yet emphasize the reliance on technology to extend human vision~\mbox{\citep{vertesi2021seeing,  kessler2012picturing}}. 
	Looking forward, ngEHT might also find it valuable to seek models beyond the history of scientific images when making decisions on how to represent data. This could include  representation of movement in film or video games, or examining the work of artists who use scientific data as the basis of their aesthetic explorations~\citep{Clarke}. EHT images of M87* and Sgr~A* have elicited a range of responses (from awe to disappointment) and have already shaped  the iconography of black holes~\citep{skulberg2021}. ngEHT imaging represents an opportunity to consider once again how imaging decisions, whether motivated by scientific or aesthetic concerns, shape the scientific and public perception of black holes.
	
	\section{Foundations}\label{foundations}

\textbf{Coordinating author: Elder, J.; Contributing authors: Ashtekar, A.; Doboszewski, J.; Enander, J.; Lesourd, M.; {Murgueitio Ramírez, S.}; Schneider, M.D.; Thresher, A.C. and Weatherall, J.O. }

	
	\subsection{Introduction}\label{intro} 
	The Foundations focus group is an extension of the existing Foundations Seminar at the Black Hole Initiative (BHI). This seminar provides a venue for discussion of foundational issues relating to black holes. 
	Previous themes of the seminar include: singularities, black hole thermodynamics, the analytic extension of the exterior Kerr metric, and theory vs.\ observation in astrophysics (among others). As we take on a new role as a focus group of the HPC working group, we will aim to facilitate further discussion of these themes in the context of the ngEHT. 
	
	In what follows, we illustrate issues that arise from such discussions. To do so, we narrow the focus to the final theme in the above list: bridging the gap between theory and observation. In Section \ref{challenges}, we provide some examples of where challenges arise for the applicability of theoretical results to real-world black holes. This includes a discussion of the no-hair theorems in Section \ref{no-hair} and a discussion of the relationship between concepts like mass, charge, and angular momentum in cosmological settings with and without a (positive) cosmological constant, in Section \ref{lambda}. Then, in Section \ref{theory_observation}, we sketch some philosophical responses to these apparent challenges. 
	
	The key questions that we seek to address in this section are these: how do we (or should we) apply formal mathematical results to a messy world where many of the assumptions behind those results are not, strictly speaking, realized? Furthermore, how can empirical results be brought to bear on theory in such cases? Our goal is to address such questions in the context of (supermassive) black holes such as those observed by the EHT and ngEHT. 
	While the discussion of these questions here is only a beginning, answering such questions in the future will have important consequences for our understanding of applications (and tests) of theoretical results using the ngEHT array. 
	
	Overall, this section serves as an example of the kinds of discussion that will continue to take place within the Foundations seminar as it takes on a second, complementary role as a focus group within the HPC working group. In addition to the theme discussed here, singular spacetimes, black hole thermodynamics, and other foundational topics concerning black holes will be the subject of ongoing philosophical discussion.  
	
	\subsection{Challenges for the Applicability of Theory to Astrophysical Black Holes: Two Examples}\label{challenges} 
	Astrophysicists and astronomers often refer to exact solutions of the Einstein field equations---especially the Schwarzschild and Kerr metrics---when describing and interpreting their observations but there are potential problems with this. 
	The Schwarzschild and Kerr metrics are highly idealized, involving assumptions that might not be physically realistic (see~\cite{booth2005black} for a discussion of this point in the context of the notion of an event horizon). For example, astrophysical black holes exist in the presence of matter fields, in a universe whose expansion is characterized by a positive cosmological constant, whereas these two metrics are solutions of the vacuum Einstein field equations and are asymptotically flat. It is therefore imperative to investigate the domain of applicability of these descriptions for astrophysical black holes. This means carefully explicating the ways that these solutions are used and examining the conditions under which the idealizations inherent in these solutions may or may not be problematic. 
	
	For illustrative purposes, we briefly consider two examples: the physical relevance of the no-hair theorems and the applicability of quantities such as mass, charge, and angular momentum for $\Lambda >0$, where $\Lambda$ is the cosmological constant. 
	
	\subsubsection{No-Hair Theorems}\label{no-hair}
	It is widely assumed that the geometry around astrophysical black holes is well described by the Kerr (or Kerr--Newman) family of metrics.\endnote{Or, in the context of a positive cosmological constant (see Section \ref{lambda}), perhaps instead one assumes a Kerr-de Sitter (or Kerr--Newman--de Sitter) metric. A good recent discussion of black holes with positive cosmological constant is in (\cite{Chrusciel2020gbh}, ch.5). One way to give these metrics is by writing them in Boyer-Lindquist coordinates, including some functions $\delta$ and $\sigma$, which are functions of radius, spin, mass, and $\Lambda$. The mass read off from such a solution is the same as the mass of the Kerr metric.} The justification for this is based on the application of so-called `no hair' theorems, according to which stationary black hole spacetimes solving the Einstein field equations in vacuum, or the Maxwell--Einstein field equations with an electromagnetic stress-energy tensor, are exhausted by the Kerr and Kerr--Newman families of metrics, respectively. 
	
	However, this line of reasoning depends on a range of assumptions that may be called into question for physically realistic black holes. 
	First, the no-hair theorems apply to stationary black holes (see Section 1 of~\cite{chrusciel2012stationary}), so their application relies on the assumption that astrophysical black holes eventually settle down to a stationary state.
	Second, existing no-hair theorems rely on various mathematical assumptions that are highly unrealistic. In the standard formulation, analyticity of the spacetime metric is required in order to show the existence of the appropriate Killing vector fields; but astrophysical modeling of gas and plasma strongly suggests the presence of shocks in the vicinity of a black hole, making analyticity an implausible assumption. 
	Third, the no-hair theorems are known to fail in the presence of matter fields (other than electric fields); see Section~5 of~\citep{chrusciel2012stationary} for a variety of examples arising if the source side of the Einstein's field equations is a (classical) Yang--Mills term.
	
	This illustrates some important concerns about the applicability of no-hair theorems for astrophysical black holes. Given that several of the assumptions behind the theorem do not, strictly speaking, hold in reality, to what extent should we expect real black holes to be well-described by the Kerr(--Newman) metrics? Furthermore, are there ngEHT observations that might provide evidence of deviations from Kerr(--Newman)? 
	No such deviations have been observed by the EHT to date. However,~ref.~\cite{EHT2022_paper6} provides constraints on potential deviations from the Kerr metric based on the 2017 observations of Sgr~A*.  
	
	\subsubsection{Mass, Charge, and Angular Momentum in \texorpdfstring{$\Lambda > 0$}{Lambda > 0}}\label{lambda}

	
	If we study the Einstein field equations with $\Lambda=0$, adopting certain assumptions about global spacetime structure (e.g., that the underlying manifold is simply connected at infinity and spacetime geometry is asymptotic to Minkowski spacetime), the theory of general relativity seems to single out a small number of global quantities---ADM mass, charge, and angular momentum\endnote{ADM stands for Richard Arnowitt, Stanley Deser and Charles W. Misner, authors of the Hamiltonian formulation of general relativity known as the ADM formalism, within which the ADM quantities are defined.}---which play a central role in understanding and quantifying basic astrophysical phenomena. However, cosmological observations support the conclusion that the accelerated expansion of the universe is well described by a positive cosmological constant, i.e., $\Lambda>0$~\cite{Planck2020}. Some have taken this empirical finding to signify the need for a better understanding of the character of global quantities in an asymptotically de Sitter universe, to replace the ones currently in use (see~\cite{schneider2022emptyspace} for discussion on this inference, including caveats). Recent progress in defining and understanding counterparts of the ADM quantities in the $\Lambda>0$ case has been made by Abhay Ashtekar and collaborators~\mbox{\cite{Ashtekar-Bahrami2019no-incoming_raditation,Ashtekar-etal2014framework,Ashtekar-etal2016gws_isolated_systems,Ashtekar2017Implications}}. Unlike the familiar ADM quantities noted above, these new ones take for granted different assumptions about global spacetime structure.
	
	It would be prudent to clarify the relationships between the global quantities in the $\Lambda=0$ and $\Lambda>0$ cases, including the role of the flat case in characterizing black holes in the $\Lambda>0$ context. In particular, doing so seems necessary to interpret what astronomers are observing when they measure the mass, spin, etc.\ of real astrophysical black holes under an idealizing assumption that the cosmological constant vanishes. The central issue here is a general question about how to interpret global properties and asymptotic spacetime assumptions as relevant to astrophysical modeling. However, a further issue arises: how to interpret specific asymptotic assumptions within idealized models in a situation where physical expectations about the expected asymptotic spacetime structure ($\Lambda>0$) are very different from that of an idealized model? One might hope that a systematic understanding of isolated systems includes an interpretation of asymptotic spacetime assumptions as becoming approximately true `far away’~\cite{Wallacepittphilsci19729}. However, in a $\Lambda>0$ context, there would seem to be such a thing as moving `too far away' from the isolated system under study (due to the presence of cosmological horizons that separate distant observers from the system). Therefore, considering the particular case of $\Lambda>0$ complicates any such story about asymptotically flat structure becoming approximately true.

	\subsection{Theory and Observation: Bridging the Gap}\label{theory_observation}
	
	Purported problems like the examples above elicit a range of responses from strict to pragmatic. One guiding question for the Foundations group moving forward is the following: how can we apply theory to observations (and vice versa) when strictly speaking there is a mismatch (e.g., the conditions of theorem are not met in the real world)? Furthermore, how can this be justified? Answering these questions will generally be sensitive to the details of the case---including the precision of the description needed and the stability of the theoretical results across changes in assumptions. 
	For now, we defer detailed consideration of the above examples to future work. Here, we instead outline some different perspectives on the general theme along with some guiding philosophical morals. 
	
	A strict approach, prioritizing mathematical rigor, is to adopt a cautious stance and not apply concepts or models outside the domain in which the assumptions behind them are true. If the assumptions behind a theorem are not true then it is not considered to be physically relevant. This approach embodies a conservatism toward epistemic risk, prioritizing the avoidance of errors over pursuing potentially fruitful (but risky) avenues of reasoning. On such a view, the issues described above are indeed considered to be problematic, amounting to a pressing need for further study and understanding. The no-hair theorem case, for example, suggests a need for a better understanding of black holes beyond the Kerr--Newman family. The manifest failure of no-hair theorems in the presence of matter fields means that we should absolutely expect to see deviations from the Kerr metric in the near-horizon regime. A better understanding of what these deviations could be and how we might test for them should be part of the scientific landscape.  
	
	A more pragmatic approach to these issues is based on a different conception of the roles of models in scientific inferences. Indeed, a recent `pragmatic turn'~\cite{Bokulich-Parker2021} in the philosophy of modeling and measurement has led to a greater emphasis on epistemic goods such as reliability (e.g.,~\cite{Cartwright-etal_Tangle}) and adequacy for purpose (e.g.,~\cite{Parker2020}) over truth. 
	
	For Cartwright et al.~\cite{Cartwright-etal_Tangle} the mismatch between models and the real world is resolved by noting that science gets to truth via reliability. Indeed, the vast majority of science is not the kind of thing that takes truth values. Models, along with things like measures, experiments, codes, narratives and techniques are essential parts of science; and yet, what would it mean to say a code or technique is true? Instead, we can ask the more important question---are they reliable? If so, for what? In what context? On this view, reliability, far more than truth, captures the actual goals and structures of science and helps explain why models are useful for black hole physics---because our goal is to create reliable systems for capturing black hole dynamics and properties: systems that, in turn, give us reliable results for the particular job at hand. We have only to look at processes of model-building and model-selection to see this in action---particular idealizations are chosen, and values set, that get us closest to useful results. This, in turn, is taken to be a proxy for truth that works provided we remain within the context the model is built or adapted to be useful for.\endnote{This perspective also has implications for how we think about the use of robustness reasoning discussed in Section \ref{robustness}.}
	
	From this perspective, asking whether the assumptions underlying various foundational results are true is misguided, and better questions would be whether the assumptions are reasonably clear and the results are useful for various purposes. This seems to be the attitude adopted by many working physicists.  However, even if one grants that a pragmatic, or even instrumentalist, attitude to foundational issues is justified for many practical purposes, one might think simply dismissing the foundational worries raised above is too fast. One reason is that a way in which our models can be useful and even reliable is by identifying points of tension in our understanding of a given physical system---in this case, black holes. Those points of tension, where models with apparently overlapping domains disagree, or where it is unclear whether the assumptions of this or that theorem truly apply to a given case, have historically been catalysts for developing new physics that can explain why different, apparently inconsistent, models nonetheless work in different contexts.  A too-radical form of instrumentalism about scientific modeling would presumably reject the demand to make our models consistent, or to at least resolve the tensions that may arise between them~\citep{stein,mitsch}.      
	
	Future ngEHT observations will play a mediating role, bridging the gap between real astrophysical black holes and idealized theoretical descriptions of them. Doing so will mean scrutinizing the reliability of our best models of black holes and the domains of applicability of theoretical results pertaining to them.      

	\section{Collaborations}\label{collaboration}

\textbf{Coordinating author: Martens, N.C.M.; Contributing authors: Doboszewski, J.; Elder, J.; Galison, P.; Lalli, R.; Marcoci, A.; Nguyen, J.; Ritson, S.; Schneider, M.D.; Skulberg, E.; Sorgner, H.; Van Dongen, J.; Wu, J. and W\"{u}thrich, A.}
	
	\subsection{Introduction} \label{collabintro}

	The Collaborations focus group lies at the intersection of various approaches within the humanities and social sciences, including history, philosophy, sociology, science and technology studies, integrated history and philosophy of science, and law. As a result, we combine a mix of different methodologies, including literature analysis, comparative case studies (e.g., ATLAS, LIGO-Virgo, IPCC, Hubble, JWST; see below), tools from the digital humanities, interviews and surveys. This will enhance our ability to engage with the rest of the ngEHT collaboration in a way that includes a diversity of opinions, all with the aim of supporting a constant dialogue to provide real-time recommendations to the ngEHT collaboration, qua collaboration, at each of its various stages of development and operation.
	
	The focus group concerns itself with the relationship between individuals and the ngEHT collaboration as a whole. To address ngEHT’s social epistemology (i.e., how knowledge is produced in social groups such as scientific collaborations), we delve into how knowledge is conditioned by the collaborative production of data, images, and text, and what the process of negotiation entails for its claims about the world.\endnote{\label{authorfn}For instance, it is well known that the more authors a scientific paper has, the more conservative the claims in the paper may be, and the longer (on average) the paper, as well as its title, tend to be~\citep{lewisonhartley2005}. Single-authored blogs tend to be more readable than blogs authored by two authors, as measured by the Flesch readability score, despite no difference in average sentence length~\citep{hartleycabanac2016}. If this can be extrapolated to journal papers with large numbers of authors, the ngEHT may want to experiment with breaking up papers into separate papers, each of which is written by a smaller set of authors, and/or for the writing to be done by the smallest possible number of people with other members of the project providing input in other ways/at other stages (e.g., everyone is involved in outlining the structure of the paper and the eventual editing, but not in the writing process in between). The latest report by the Intergovernmental Panel on Climate Change (IPCC) provides a model of such a practice. A first draft by one of their working groups (WG1) was written by just the working group, comprising 240 scientists (Assessment Report [AR] 1 WG1 IPCC, 2021). After this, a much larger number of scientists from around the world provided comments that were incorporated into subsequent drafts. The ngEHT could consider writing papers following this model, scaled down according to the smaller number of scientists involved. } 
It is clear from previous large-scale collaborations---and the ngEHT will be no exception---that \emph{the establishment of fact,} 
and \emph{what constitutes a fact} are to some extent the result of the social negotiation of consensus.\endnote{On the historical contingency of our notion of fact, see~\citep{daston2001,poovey1998,tenhagen2019,dewaal2020}.}  Thus, group structure and the distribution of authority play a direct role in what counts as knowledge. For instance,  the particle physics community has converged on near-universal conventions regarding the determination of facts---five sigmas are required for a discovery---whereas only two sigmas are required to exclude new physics hypotheses.\endnote{On the role of 'sigma's' in modern physics, see~\citep{franklin2013}.} In contrast, there is currently no such shared standard in astrophysics. 

The importance of human judgments was evident throughout the EHT imaging process. For example, multiple imaging pipelines produced their own images of M87* based on a range of choices (imaging methods, specific algorithms, priors and other inputs, etc.). These results then had to be aggregated in order to present a result that represented the collective judgment of the collaboration. The averaged image of M87* released in 2019 reflects a particular choice about how to do this aggregation (cf., Section \ref{AIV}).\endnote{On the practice of averaging over black hole images as epistemic practice, see~\citep{philosophyoftheshadow,theedgeofallweknow}.} 

\subsection{\label{subsecknowledgeformation}Knowledge Formation and Governance: Top-Down vs.\ Bottom-Up}

Large-scale scientific collaboration can take place within a variety of governance/ organizational structures, ranging from top-down hierarchical structures to more loosely organized bottom-up collaboration in the absence of a formal governing structure. 
We, the ngEHT collaboration, see ourselves as (ideally) being located somewhere in the middle of this spectrum---in particular somewhat closer to the bottom-up extreme than the EHT collaboration. In this section, we briefly illustrate this claim by contrasting the ngEHT with instances of scientific collaboration found at either extreme---specifically, the particle physics collaborations ATLAS and CMS associated with the Large Hadron Collider (LHC), the gravitational-wave-detecting LIGO--Virgo collaboration (LVC), and the Intergovernmental Panel on Climate Change (IPCC).  

At the top-down extreme of the spectrum, partially exemplified by ATLAS and CMS, as well as the LVC, we find hierarchical structures with a centralized, physical headquarters and funding stream, with one or a few main instruments or purposes, and a large number of committees that decide which collaboration papers are published and how, which members get to present at which conference, etc. The collaboration is prioritized over the individual member; consensus is prioritized over dissent and diversity of opinions, with dissent being procedurally dealt with internally before (consensus) results are published. This structure facilitates a strong group identity, obtaining a large amount of funding for a dedicated, coordinated purpose, and achieving that purpose in the most efficient way possible. However, there is a risk that individual credit and creativity are lost to some extent. 
In contrast to these top-down examples, the ngEHT is a loosely organized, informally scripted, yet formally documented collaboration. Although workshops and conferences bring together researchers for short periods of time, observations will take place from different continents, researchers usually work from different geographical locations, and no building has been constructed for the purpose of housing ngEHT research. Instead, the asynchronous electronic infrastructure~(\citep[]{Galison:1999ad}, p. 159) of Overleaf, Slack, Google documents, slides, and telecons will be used to coordinate matters.

At the bottom-up end of the spectrum, we do not find formal collaborations per se but instead entire scientific communities with a common subject and a more or less uniform research culture. In such cases, authors coalesce in and out of projects, with members of the community communicating via conferences and peer-reviewed publications rather than in a physical headquarters. In this bottom-up model, individual groups can pursue any research direction that they themselves consider fruitful---as long as they manage to get funding---and publish dissenting results. A coherent, negotiated narrative connecting all these results and delineating the \emph{facts} is more likely to be established later (if at all), through review papers and review presentations in textbooks. Particularly striking examples are meta-analyses in medical communities or the recent report by the Intergovernmental Panel on Climate Change (endnote \ref{authorfn}) which synthesizes 14,000 papers from the climate science community. 
In contrast to such extreme bottom-up examples, some sustained collaboration is required to achieve the ngEHT's main goals: financing and building additional telescopes and coordinating the whole network of telescopes so that it has access to; the joint reduction of data; and, finally, reporting its findings in publications. Moreover, it is important to stress that maximizing the benefits of bottom-up approaches does not come for free; it is not a mere matter of the absence of a top-down governance structure, but also the implementation of positive measures that bring out the advantages of bottom-up approaches, such as room for diversity and individual creativity. 

One important challenge for the ngEHT then, regarding the spectrum of bottom-up versus top-down approaches to social epistemology and governance, is to be the best rather than the worst of both worlds. In the remainder of this section, we outline some preliminary thoughts on how this can be achieved. In particular, we discuss the need to facilitate dissent (Section \ref{socepi}) and to adopt a governance charter (Section \ref{governance}).

\subsection{Knowledge Formation: Differences of Opinion} \label{socepi}

Should large scientific collaborations aim for consensus? The extent to which consensus is ideal for a scientific collaboration depends on how consensus is construed. First, we can consider the \textit{unit} of consensus: should the group agree on individual propositions or collections of (logically connected) propositions?\endnote{Work in judgment aggregation theory highlights the impact these relations can have on the consistency of the group attitude, see~\citep{list12}.} Second, we can consider the \textit{bearer} of consensus. In the first instance, whether a group should aim at consensus may depend on the nature of the collaboration: what ties the individuals together?\endnote{See~\citep{bird14}'s distinction between the ``commitment'' and ``distributed'' models of group knowledge.} In the second instance, when we attribute consensus to a group, are we ``summarising'' the attitudes of the individuals, or does the collaborative aspect add something to this---possibly in the sense of a ``plural subject'' or a ``group agent''~\citep{quinton76, gilbert89,listandpettit11}? Third, we can distinguish between at least two \textit{attitudes} relevant to the consensus: if a group is in consensus does it (or each member of it) hold a consensual belief, or a consensual acceptance, where different epistemic norms are associated with each attitude (e.g., belief requires a commitment to truth while acceptance may not)~\citep{wray01, gilbertandpilchman14, Dangandbright2021, dethier22}.\endnote{The distinction between belief and acceptance can also help us conceptualize the role of idealization in science, as discussed in Section~\ref{theory_observation}, see, for instance,~\citep{elgin2017}.} Fourth, we can ask about the \textit{extent} of consensus: at one extreme consensus might be identified with unanimity, but some level of dissent may be consistent with consensus, and indeed, as we discuss below, even encouraged~\citep{dellsen21}. Clarifying each of these dimensions allows us to ask more fine-grained questions about the nature and desirability of consensus (e.g., we can attribute a consensus belief to the group without necessarily requiring that all, or even any, of the individuals, believe all, or even any, of the propositions the group believes, although they may accept them in virtue of being in the~collaboration).

The above requires us to take a step back and ask what \emph{being in the ngEHT collaboration} 
actually means. Issues such as who may be a member of the ngEHT collaboration and an author of the collaboration's papers need to be made explicit. How is membership established, and what does it imply to be a member? Which rights, responsibilities and credits follow from membership? Who may become a member? Is a vetting procedure required, and which members get to decide who else may become a member? Should there be different types of membership?  Are all members also on the author list of collaboration papers? Is it possible to be a member without being an author? Might different types of authorship (e.g., data compiler, data analyst, text writer) be desirable? How are papers written and what epistemic goals might be favored by such a process? What happens if the collaboration is succeeded by another, or splits up: who owns the collaborative knowledge? Answers to these questions make clear who is a party to making knowledge; and thus also, what constitutes knowledge.\endnote{Compare, e.g.,\ with discussions on including string theorists as physicists~\citep{galison2004, vandongen2021, greenspan2022}.} In the near future we aim to survey how different modalities of membership and authorship have been crafted in comparable yet different collaborations (ATLAS at the LHC, LVC, and IPCC), and make an inventory of current practices in the EHT and ngEHT collaborations, including an analysis of their advantages and drawbacks. 


Returning to consensus, some construal of consensus is  \emph{prima facie} valuable  and to be expected in scientific collaborations. First, because epistemic peers presented with the same evidence are, on the first approach, expected to reach the same conclusions~\citep{hulme2013lessons}.
Second, the higher the number of independent and competent scientists who believe a particular claim, the more likely it is to be true (the relevant result is a generalized Condorcet's Jury Theorem~\citep{listgoodin2001,sep-jury-theorems}). 
Third, the stronger the consensus for a claim, the more likely it is for the general public to accept it~\citep{lewandowsky2013pivotal}. Finally, a lack of consensus is often what politicians and lobbyists use to undermine the findings of scientific collaborations~\citep{oreskes2011merchants}.

On the other hand, there are reasons to be wary of some construals of consensus.
Consensus between individuals may be impossible to achieve in contexts where the collaboration involves individuals with different values and/or disparate areas of expertise. Furthermore, the fact that epistemic peers \emph{may} reasonably disagree on substantive issues motivates the applicability of judgment aggregation theory to scientific collaborations~\citep{bdh17, marcociandnguyen19}.
Finally, when consensus is enforced through a collaboration's policies in a top-down fashion (cf.~Section~\ref{subsecknowledgeformation}), this may disincentivize deliberation and the exploration of competing hypotheses~\citep{beatty_moore_2010}; it may also produce the appearance of agreement when there is none~\citep{Fuller1986,hulme2013lessons,weatherall2021conformity,Fazelpour2022-FAZDTA}. 

It is thus important to find a good balance between top-down and bottom-up approaches to structuring an organization (cf.~Section~\ref{subsecknowledgeformation}) that promotes consensus-building without prematurely suppressing dissent.
Having a diversity of beliefs and practices among team members can be epistemically beneficial to science.
For example, individuals in collaboration may draw on different (and possibly even competing) sources of evidence and theories in order to justify their conclusions~\citep{Dang2019}.
Moreover, if all team members test the same hypothesis (and especially, by means of the same methods), they may prematurely settle for false beliefs. Several authors (notably~\citep{zollman2010epistemic}) have advocated for a period of \textit{transient diversity} during scientific research when different epistemic options are sufficiently tested before the community settles on a consensus. 

Mechanisms that allow for or encourage transient diversity thus present strategies to promote a desired kind of creativity at the group level within bottom-up research contexts~\cite{currie2019existential,schneider2021creativity}. While the influence of (diverging) non-cognitive values in science is unavoidable, it is not necessarily pernicious~\citep{parker2014values}, and transient diversity could provide one such mechanism. Indeed, a more inclusive representation of values and perspectives is expected to produce epistemically more robust results~\citep{longino1990science}. Increasing transient epistemic diversity may also be helped by incorporating perspectives from marginalized groups into the scientific inquiry~\citep{Wylie2003-WYLWSM,mills2007white,du2008souls,wu2021}. Furthermore, facilitating minority views and carefully publicizing (partial) dissent increases transparency and enhances rather than erodes the credibility of the collaborations’ conclusions~\citep{hulme2013lessons,dellsen21}. 
One motivation for this bottom-up line of reasoning stems from the social turn in the philosophy of science~\citep{longino2018fate}: emphasizing the political, social, and psychological aspects of scientific collaborations encourages the idea that trustworthy decisions in science, as in other social institutions, requires deliberation, transparency and openness.
Enforcing consensus goes against these~norms. 

In light of the above, what techniques and policies should guide collaboration within the ngEHT? 
Firstly, there are several mechanisms that can generate (transient) diversity. Of particular interest are modeling results~\citep{pittphilsci19428} showing that the less connected the epistemic community is, the more likely it is to converge to the true belief---but the slower it is at doing so~\citep{zollman2007communication,zollman2010epistemic,lazer2007network,fang2010balancing}. For high stakes frontier research where it is important to be correct, it may be warranted to temporarily limit communication between team members. For instance, the limited communications between the imaging teams at the EHT may have epistemically benefited the final results~\citep{galison2021interview,pittphilsci19428}.\endnote{Interesting in this regard is the current ngEHT analysis challenge, where part of the collaboration creates a training set from simulated signals with noise added to them (and potentially also some fake signals), with another part of the collaboration honing their analysis tools on this training data without knowing how it was created.}

Moreover, there already is evidence regarding the benefits of including groups traditionally excluded from knowledge production; some local and Indigenous communities on EHT’s sites would have relevant scientific knowledge that other team members do not (see~\citep{Wylie2020-WYLCII} for collaboration with Indigenous communities). However, empirical and simulation results show that marginalized and minoritized people often receive less credit in scientific collaborations~\citep{ross2022women,sarsons2021gender,rubin2018discrimination}. Given this, collaborative teams should consider explicit, ongoing discussions about credit assignment procedures, being particularly vigilant about assigning fair credit to marginalized knowers' contributions---this will be one of the roles of the ethics committee proposed in the next subsection.
A related concern is that creative research is stifled and individuals are prevented from developing diverse and novel ideas. Large research collaborations may tend towards conservatism, in part stemming from multiple requirements for collective approval~\citep{ritsonforth} and a preference for well-tested over novel approaches~\citep{merz2022organizational}. When considering how we might ideally organize a research collaboration, it is thus important to consider creativity from both an individual and a collective perspective~\citep{ritson2021creativity}, including the opportunities for researchers to publish individual contributions to collaborative research, such as PhD theses~\citep{sorgner2022constructing}.

The Collaborations focus group will also explore how ngEHT members interact with one another. 
Methodologically, we can use concepts and tools from network theory to quantitatively investigate the structure of the collaboration and its change over time. By using a multi-layered network perspective of socio-epistemic networks we can investigate how the social structure is related to the production of new knowledge~\citep{lalli_dynamics_2020, lalli_socio-epistemic_2020}. Network approaches also allow us to understand the flow of information within the collaboration. An illustrative example in this regard is recent work analyzing more than 20,000 emails sent via internal mailing lists of {a major particle physics collaboration}~\citep{wuethrich2022}. 
This analysis revealed a pronounced sub-structure of the communication network featuring smaller ``communities'' within the collaboration. The communication network is also relatively dense and, in a network-theoretical sense, less hierarchical than most such networks, which is surprising given the top-down governance structures in place. Such analyses of communications networks may provide insight into how large-scale collaborations collectively produce~knowledge.

Similar network analyses could also be done for the ngEHT. This descriptive project could also inform the normative guidance that we provide to the collaboration; the analyses could be used to test hypotheses about what communication structures might be particularly conducive to epistemic success, and which mechanisms and governance structures would foster such communication. 
This work could then be connected to the rich body of literature spanning decision theory, social psychology, and mathematics that explores the advantages and drawbacks of different ways of structuring deliberation between, and eliciting judgments from, experts~\citep{Morgan2014,burgman2016trusting}, as well as formal frameworks for conceptualizing the relationship between the attitudes of individuals and the attitudes of the group~\citep{list12}.
For the ngEHT, the exact balance between seeking collectivist consensus from the outset or operating via integration and trade-offs between autonomous viewpoints will depend on how data and responsibilities are shared among members, whether there are distinct organizational sub-units within the collaboration, and what the final arbiter is in cases of conflict (e.g., whether an appeal to a higher authority is possible, and how that authority is legitimized). The authorship of publications (whether they are mainly collectively authored or authored by distinct groups within the collaboration) will likely reflect these organizational norms~\citep{vertesi2020shaping}.

In sum, it is clear that it would be beneficial for the ngEHT not to enforce consensus in the top-down fashion known from, among others, the various LHC collaborations. The Collaborations focus group aims to enrich the somewhat abstract existing literature by investigating concrete mechanisms and organizational structures that can maximize the benefits of epistemic diversity, applicable to the ngEHT context via a detailed analysis of the practice of the ngEHT collaboration with tools from the digital humanities and with internal surveys. It is crucial that these organizational structures are geared towards representation, diversity, sufficient freedom for individual creativity, the appropriate balance between transparency and epistemic distance at various stages of the collaboration, and appropriate assignment of credit, as elaborated upon in the following subsection.

\subsection{Governance} \label{governance}





Well-structured governance is key to the future of collaboration. A main task for the Collaborations focus group will be to systematically analyze the organizational structure of various similar collective entities, including LIGO-Virgo, EHT, ATLAS, CMS, CERN, IPCC, the UN, Hubble and JWST, to identify their main benefits and drawbacks. Surveys conducted among ngEHT members, based on a similar survey conducted within the EHT collaboration, will also provide valuable data moving forward. These lessons will be synthesized into the optimal governance model for the ngEHT, keeping in mind the desiderata and worries described in the previous subsections. 

To give the reader a tentative first impression of what such a governance model might look like, we sketch here an initial suggestion. We view this as the beginning of an ongoing conversation about the optimal governance model for the ngEHT collaboration. 
This model will then be iteratively tested and improved, especially with regards to how it facilitates knowledge formation and adapted as circumstances change. Given that the nascent ngEHT collaboration has already begun to take shape, it is crucial that this group make what recommendations we can---however preliminary---at this early stage. We are now in a position to influence organizational structures that may become increasingly entrenched as the ngEHT project gains momentum. 

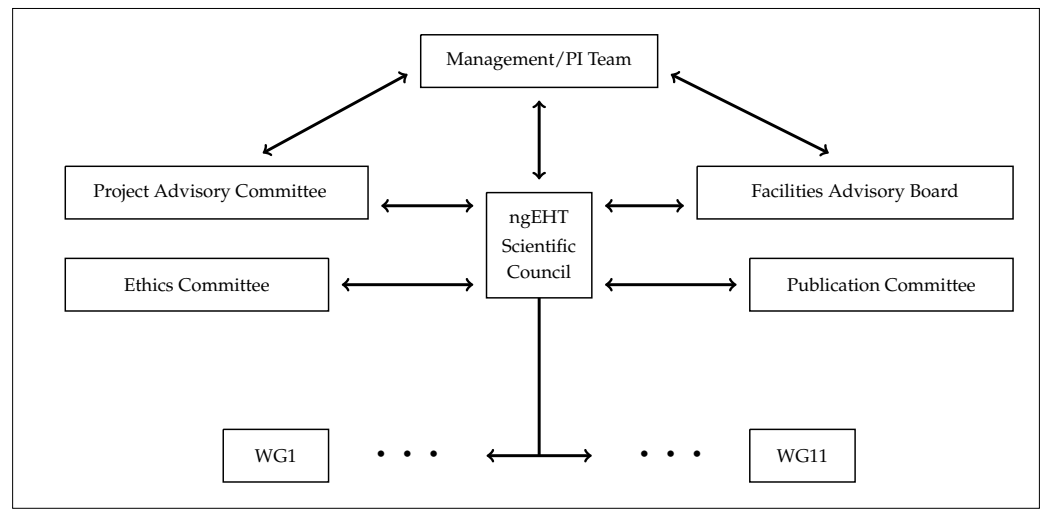
\begin{figure}[H]
	\centering
\scalebox{0.7}{    
\begin{tikzpicture}
	
	\filldraw[fill=white, draw=black] (-4,6) rectangle (15.5,15.5);
	
	\draw[black, thick] (3.75,15) -- (8.25,15) -- (8.25,14) -- (3.75,14) -- (3.75,15);
	\node [black] at (6,14.5) {Management/PI Team};
	
	\draw[black, thick] (5,12) -- (7,12) -- (7,10) -- (5,10) -- (5,12);
	\node [black] at (6,11.5) {ngEHT};
	\node [black] at (6,11) {Scientific};
	\node [black] at (6,10.5) {Council};
	
	\draw[black, thick] (0,7.5) -- (2,7.5) -- (2,6.5) -- (0,6.5) -- (0,7.5);
	\node [black] at (1,7) {WG1};
	
	\draw[black, thick] (10,7.5) -- (12,7.5) -- (12,6.5) -- (10,6.5) -- (10,7.5);
	\node [black] at (11,7) {WG11};
	
	\node [black] at (3,7) {\textbullet};
	\node [black] at (3.5,7) {\textbullet};
	\node [black] at (4,7) {\textbullet};
	
	\node [black] at (8,7) {\textbullet};
	\node [black] at (8.5,7) {\textbullet};
	\node [black] at (9,7) {\textbullet};
	
	\draw [-, ultra thick] (6, 10) -- (6,7);
	\draw [<->, ultra thick] (5, 7) -- (7,7);
	
	\draw [<->, ultra thick] (6, 13.75) -- (6,12.25);
	
	\draw [<->, ultra thick] (4.75, 10.25) -- (2.25,10.25);
	
	\draw [<->, ultra thick] (7.25, 10.25) -- (9.75,10.25);
	
	\draw[black, thick] (10,10.75) -- (15,10.75) -- (15,9.75) -- (10,9.75) -- (10,10.75);
	\node [black] at (12.5,10.25) {Publication Committee};
	
	\draw[black, thick] (2,10.75) -- (-3,10.75) -- (-3,9.75) -- (2,9.75) -- (2,10.75);
	\node [black] at (-.5,10.25) {Ethics Committee};
	
	\draw[black, thick] (-3,12.5) -- (2.75,12.5) -- (2.75,11.5) -- (-3,11.5) -- (-3,12.5);
	\node [black] at (-.25,12) {Project Advisory Committee};
	
	\draw[black, thick] (9,12.5) -- (15,12.5) -- (15,11.5) -- (9,11.5) -- (9,12.5);
	\node [black] at (12,12) {Facilities Advisory Board};
	
	\draw [<->, ultra thick] (3.5,14.25) -- (.75,12.75);
	
	\draw [<->, ultra thick] (8.5,14.25) -- (11.5,12.75);
	
	\draw [<->, ultra thick] (7.25, 11.75) -- (8.75,11.75);
	
	\draw [<->, ultra thick] (3, 11.75) -- (4.75,11.75);
	
\end{tikzpicture}
}
\caption{Tentative governance structure of the ngEHT collaboration.}
\label{wiringdiagram}
\end{figure}

The core of the collaboration is its eleven working groups---eight science working groups (including HPC) and three technical working groups. In other collaborations, working groups have worked particularly well to generate a sense of community and strong science. The major Principal Investigators that lead the working groups alongside (and overlapping with) the Management Team---including the ngEHT director, chief scientist, and chief engineer---take on the dual responsibilities of fiscal probity (fulfilling the contracts) and keeping a steady hand on the tiller to keep the collaboration in line with its founding goals. They would be guided and supported by a small number of governance structures (Figure \ref{wiringdiagram}): a central Scientific Council, a Project Advisory Committee, a Facilities Advisory Board, an Ethics Committee and a Publication Committee. These structures are not intended to provide top-down constraints by appointees, such as forcing consensus, but are instead (partially) elected, representative bodies that streamline the collaboration in a way that celebrates diversity and raises ethical scientific comportment to a primary aim.  

\vspace{-4pt}

\textbf{Scientific Council \& Project Advisory Committee.} 
\textls[-5]{The ngEHT, like LIGO, includes multiple sites and dozens of scientific groups. To run its program, LIGO established a scientific council (LSC) that determines the scientific priorities and the overall mission---responsible} too for science, instrumentation, communication, and operation. Composing the LSC are representatives of the various groups, in proportion to their membership size.\endnote{On the LIGO Scientific Collaboration Charter \cite{LIGO2020}.} The ngEHT might follow a modified version of that model which offers a way for the membership to shape scientific and technological policy and to facilitate decisions about priorities (such as targets, observation cadence, instrument standards, and aims). The ngEHT Scientific Council would be composed of representatives chosen by the constituent groups---no such representative body exists within the EHT collaboration. Where participating institutions or other stakeholders, including local communities and junior members, are too small to field separate representatives, they could be grouped together to form a larger body. The elected council would receive advice from the already existing Project Advisory Committee/Science Advisory Board), consisting of appointed, experienced and mostly external scholars, including Nobel laureates. 

\textbf{Ethics Committee \& Transparent Ethical Charter.}  In founding the ngEHT, a charter specifying structure is desired, but should equally include transparent record keeping, voting procedures, and appointments as well as principles of membership, publication, authorship, credit, and conflict resolution. Along with these procedures, the charter would lay down a guiding, forceful commitment to diversity, equity and inclusion, as well as to ethical comportment regarding fairness, respectful interactions, and accountability. Putting this in the founding charter would give it the weight it deserves, showing these values are foundational, not \emph{pro forma}. As groups join the ngEHT, it would be essential, in addition, to have a Memorandum of Understanding underscoring commitment to the charter and to the particular roles and responsibilities of the group. A high-level ethics committee---ideally its members would include several members of the History Philosophy Culture Working Group---would be tasked with drafting this charter to be sent to the rest of the collaboration for feedback, with overseeing the adherence to this charter once in place, and with updating the charter based on continuous feedback. It should maintain and publicize policies to promote an equitable, inclusive, and welcoming workplace. This committee could also include or run elections to identify ombudspersons and mediators as part of a broader mandate to do all in its power to stop intolerable actions visited upon collaborators such as harassment, bullying or marginalization on the basis of race, gender, nationality, or identity. 

\textbf{Facilities Advisory Board.} The ngEHT will use some established facilities and so, in part, resembles an experiment at a particular facility telescope---the ngEHT will apply for time. Essential to realizing its mission, the ngEHT aims to build approximately ten additional sites beyond the existing telescope facilities made use of by the EHT: five in a first phase with an additional five to follow. 
The Facilities Advisory Board would consist of representatives of some of the telescopes or groups of telescopes, and, if needed, scientifically-relevant facilities (e.g.,\ large-scale computation/correlators) even if they are not direct stakeholders. Note that the Facilities Advisory Board and Project Advisory Committee are separate entities, in contrast to the structure of the EHT collaboration.

\textbf{Publication Committee.} The aim of the publication committee would not be to provide negative constraints beyond standard checks regarding the use of proprietary data. It is not to be a gatekeeper that approves the official opinions and results of the members of the collaboration. Instead, its aim is positive: to streamline the process of publications through the collaboration and work of smaller subsets of members that relates to the ngEHT, by coordinating internal review in cases where this may be helpful, by ensuring that credit is given where credit is due, and by coordinating the ngEHT science book and other strategies that enhance the overall visibility of ngEHT related outputs, all in line with policies set out in the ngEHT's charter. 

The ngEHT, like the IPCC, is an overarching framework for dozens of institutes across the world, each funded in different ways. Like the IPCC, the ngEHT has working groups. In contrast to the ngEHT, the IPCC was formed by an international compact, offering not novel research but a mechanism for collective, reliable assessment of existing research---including evaluators of different career stages, genders, and geographical regions. The ngEHT could learn from the way the IPCC has honed methods of assembling expert judges to assess both scientific/technical questions and to assist in effective final write-ups of the work. Similarly, the LHC detectors ATLAS and CMS have elaborated effective (but different) means of evaluating their own work before publication, which could serve as inspiration.



In sum, a governance model like the model proposed above would serve to support the working groups and help them excel, not by providing constraints that prioritise the collaboration over the individual working group members, but in a way that streamlines their work by ensuring diverse representation of the various stakeholders.

\section{Conclusions}\label{sec5}

This white paper has presented some---but by no means all---of the plans of the History Philosophy and Culture Working Group of the ngEHT collaboration. It is unprecedented for scholars from the humanities and social sciences to be integrated into a physics collaboration of this size, from the very beginning and with the same standing within the collaboration structure as its STEM members. We would like to cordially invite other scholars from the humanities and social sciences to join us in this exciting endeavor of making the ngEHT a prime model for interdisciplinary collaboration and recording high-quality videos of a black hole together.

\vspace{6pt} 
\authorcontributions{Writing---original draft preparation, all authors; writing---review and editing, all authors; supervision, P.G.; project administration, P.G., J.D., J.E. and N.C.M.M. 
All authors have read and agreed to the published version of the manuscript.
} 


\funding{{J.~Doboszewski, J.~Elder and N.C.M.~Martens would like to thank the Volkswagen Foundation for its support in providing the funds to create the Lichtenberg Group for History and Philosophy of Physics at the University of Bonn.
M.~Gueguen and N.C.M.~Martens would like to thank the European Union's Horizon 2020 research and innovation programme for the funding received under the Marie Sk\l{}odowska-Curie grant agreements No.~101026214 and No.~101065772, respectively.
P.~Galison, J.~Doboszewski, J.~Elder, M.~Lesourd, and P.~Natarajan also acknowledge the support of the Black Hole Initiative, which is funded by grants from the John Templeton Foundation and the Gordon and Betty Moore Foundation (although the opinions expressed in this work are those of the authors and do not necessarily reflect the views of these Foundations).
J.~Elder also acknowledges the support of the ``Inductive Metaphysics'' project funded by the Deutsche Forschungsgemeinschaft (DFG), Research Unit FOR 2495 (specifically subproject B6: ``The Role of Inference to the Best Explanation in the Discovery of Gravitational Waves'').
N.C.M.~Martens, H.~Sorgner, and A.~W\"{u}thrich's contribution was made possible by funding from the DFG (FOR 2063)/FWF (I 4410-G) Research Unit ``Epistemology of the LHC'', and A.~W\"{u}thrich's contribution furthermore by funding from the European Union (ERC, Project NEPI, No.~101044932). Views and opinions expressed are however those of the authors only and do not necessarily reflect those of the European Union or the European Research Council. Neither the European Union nor the granting authority can be held responsible for~them.}
}


\dataavailability{Not applicable. 
} 

\acknowledgments{We want to recognize the early and important contributions of T. Nichols, N.~Conway (on outreach), A. Raymond, G. Fitzpatrick, M. Johnson (on technical siting) to the formation of the HPC working group, which in turn built on the already long-running BHI Foundations seminar (with many thanks e.g., to F. Azhar, M. Lesourd, E. Curiel and the participants of that seminar). In framing the scope of the still-developing Siting focus group that will report in subsequent publications, the Siting Workshop conveners and framers, including A. Thresher and P. Natarajan (later joined by D. Palumbo), thank the presenters at the first Siting Workshop which has helped guide subsequent developments: C. Prescod-Weinstein, K. Kamelamela, H. Nielson, M. Johnson, K.~Fox, J.~Havstad, T. Nichols, R. Chiaravalloti, S. Doeleman, G. Fitzpatrick, J. Houston, A. Oppenheimer.  We would like to thank Jonas Enander, Luis Reyes-Galindo, Mike Schneider \& Jeroen van Dongen for their internal review of this white paper. We are grateful for valuable discussions with the attendees of the HPC Kick-Off Workshop (Black Hole Initiative, Harvard, \mbox{Feb--March} 2021), with the attendees of the ngEHT meeting (Granada, June 2022), and with the other members of the HPC working group.

}

\conflictsofinterest{The authors declare no conflict of interest. 
} 
%


\begin{adjustwidth}{-\extralength}{0cm}			
\printendnotes[custom]			
\end{adjustwidth}

\begin{adjustwidth}{-\extralength}{0cm}

\reftitle{References}




\PublishersNote{}
\end{adjustwidth}
\end{document}